\documentclass[a4paper,11pt]{article}
\usepackage{pos}

\usepackage[british]{babel}
\usepackage{xspace}
\usepackage[output-decimal-marker={,}, binary-units = true]{siunitx}
\usepackage{chngcntr}
\counterwithin{figure}{section}
\counterwithin{table}{section}
\usepackage{subfigure}
\usepackage{cleveref}
\crefname{table}{table}{tables}
\Crefname{table}{Table}{Tables}
\crefname{figure}{figure}{figures}
\Crefname{figure}{Fig.}{Figures}

\makeatletter
\newcommand{\GeV}{\,\si{\giga\electronvolt}\xspace}
\newcommand{\TeV}{\,\si{\tera\electronvolt}\xspace}
\newcommand{\MET}{\ensuremath{\ptmiss}\xspace}
\newcommand{\ET}{\ensuremath{E_{\text{T}}}\xspace}
\newcommand{\ETmiss}{\ensuremath{\ET\hspace{-1.1em}/\kern0.45em}\xspace}
\newcommand{\pt}{\ensuremath{p_{\text{T}}}\xspace}
\newcommand{\ptmiss}{\ensuremath{\pt^\text{miss}}\xspace}
\makeatother

\newlength{\bibitemsep}
\newlength{\bibparskip}
\let\oldthebibliography\thebibliography
\renewcommand\thebibliography[1]{%
  \oldthebibliography{#1}%
  \setlength{\parskip}{\bibitemsep}%
  \setlength{\itemsep}{\bibparskip}%
}

\title{Reconstruction of jets and missing transverse momentum at the CMS experiment: Run 2 and perspective for Run 3}

\author*[a]{Andrea Malara}

\affiliation[a]{Institut f{\"u}r Experimentalphysik, Universit\"at Hamburg, Luruper Chaussee 149, Hamburg, Germany }

\emailAdd{andrea.malara@cern.ch}

\abstract{The performance of reconstruction and calibration of jets and missing transverse momentum in the CMS Collaboration, together with the latest tools for pileup mitigation, is presented. The results shown are relative to the data collected from 2016 to 2018 (Run 2) in proton-proton collisions at a centre-of-mass energy of 13 \TeV. An outlook to the new techniques foreseen for Run 3 is discussed.}

\FullConference{%
  40th International Conference on High Energy physics - ICHEP2020\\
  July 28 - August 6, 2020\\
  Prague, Czech Republic (virtual meeting)
}

\begin{document}
\maketitle

\section{Introduction}

In the hadronic environment at the LHC, many quarks and gluons are produced, which, due to QCD confinement, create a collimated spray of hadrons, which appear as a cluster of energy deposited in a localised area of the detector, called a jet.
Precise calibration of both the energy scale and resolution of jets plays a crucial role across the whole physics programme at the CMS experiment. Similarly, an accurate estimation of the missing transverse momentum (\ptmiss) is of crucial importance, for example, in standard model measurements involving the invisible decay products of the top quark, $\tau$ lepton and the W, Z and Higgs bosons, as well as in beyond the standard model searches targeting new weakly interacting neutral particles.\par
Moreover, the presence of multiple collisions in the same bunch-crossing (pileup) represents a challenge to the reconstruction and calibration procedure. Several techniques can be used to limit the effect of additional particles. An overview of the respective methods and their impact on the performance is presented in the following.

\section{Jet reconstruction}
The event reconstruction in CMS \citep{Collaboration_2008} uses a similar scheme for both the online and offline reconstruction, with some difference in the amount of information used and level of corrections applied to achieve fast performance during data-taking. Starting from a local reconstruction, the information from each sub-detector is combined and successively used by the Particle Flow (PF) \citep{PF} algorithm to identify the particles produced in each event. Jets are usually clustered, starting from PF candidates combined with pileup mitigation techniques. It is possible to reconstruct several types of jets, based on different algorithms. Furthermore, we define particle-level jets as clustered from all stable ($c\tau >1$ cm) and visible particles (excluding neutrinos) in simulated events. Finally, it is possible to calibrate the energy of the jets and propagate these corrections onto missing transverse momentum (\MET) and apply additional pileup rejection.

\section{Pileup mitigation}
\begin{figure}[htb]
\centering
\subfigure{\label{fig:Pileup}}{\includegraphics[width=0.40\textwidth]{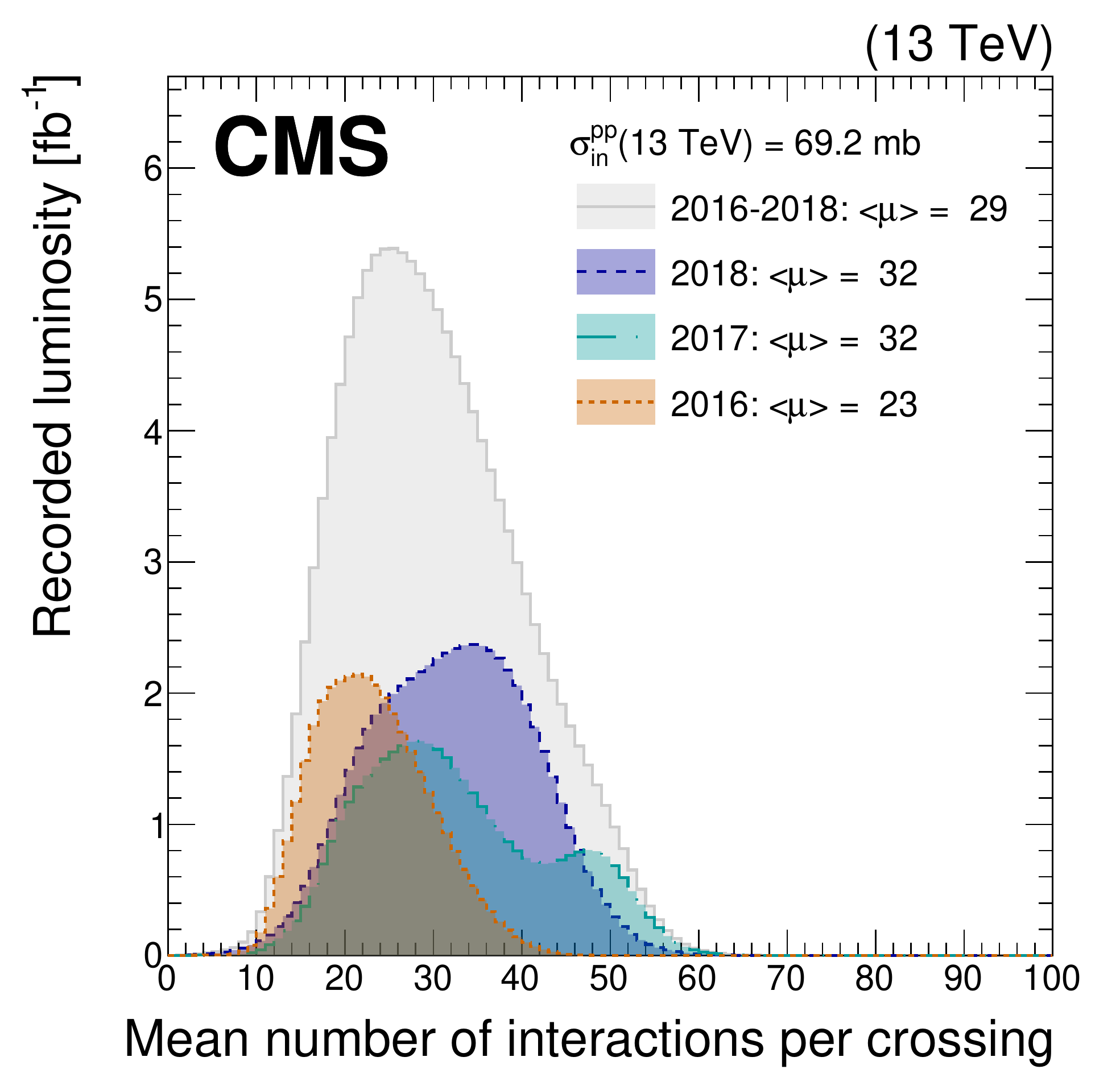}}
\subfigure{\label{fig:Puppi weights}}{\includegraphics[width=0.40\textwidth]{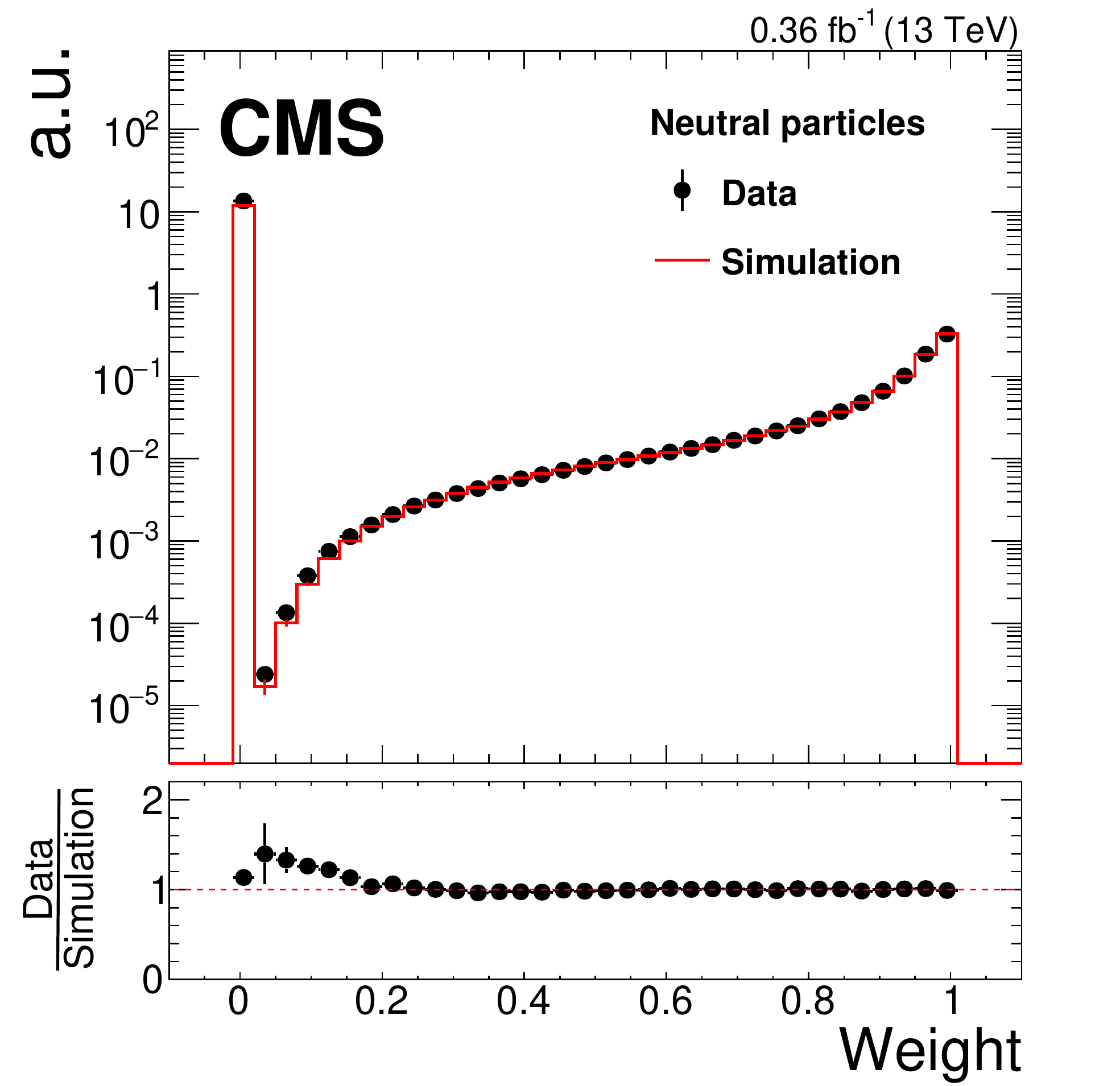}}
\caption{
Left: Distribution of the mean number of inelastic interactions per bunch crossing (pileup) in data for pp collisions \citep{Puppi}. 
Right: Data-to-simulation comparison of the PUPPI weight distribution for neutral particles \citep{Puppi}. 
}
\label{fig:LumiPileUpPuppiWeight}
\end{figure}

The instantaneous luminosities reached during data-taking imply multiple proton-proton (pp) collisions to occur in the same bunch crossing. Additional particles coming from the secondary interactions, known as pileup (PU), can deteriorate the measurement since they may be clustered in the reconstructed jets. Hence, the identification of interesting collisions has become an ever-growing challenge at the LHC. During Run 2, the CMS detector collected data with up to 60 interactions per bunch crossing, with an average pileup of 30 interactions (see \Cref{fig:LumiPileUpPuppiWeight}). Moreover, the expected average pileup for Run 3 is even higher, making PU rejection an even more challenging task.

The CMS Collaboration uses a variety of techniques for PU mitigation. One example is the Charged Hadron Subtraction (CHS) algorithm, which has been the standard method in Run 2.
It uses the information from the tracker to remove the charged particles that are associated with a pileup vertex from the jet clustering procedure.
Due to its limited coverage in $\eta$, outside the tracker no information on the charge of a particle is available; consequently, dedicated jet energy corrections are applied to account for the impact of charged PU outside the tracker coverage, and of neutral PU everywhere.

\begin{figure}[htb]
\centering
\subfigure{\label{fig:PileupSketch}}{\includegraphics[width=0.9\textwidth, trim= 10 250 0 220,  clip]{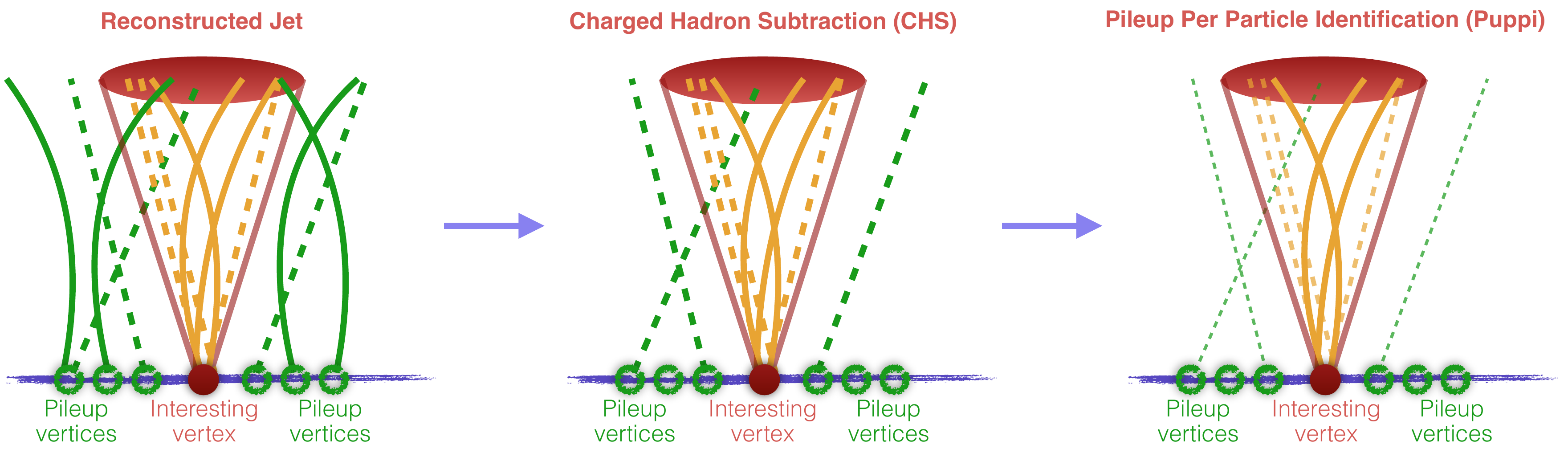}}
\caption{Sketch of PU suppression techniques. Solid (dashed) lines refer to charged (neutral) PF candidates. The weights applied by the PUPPI algorithm are represented by thin lines.}
\label{fig:PileupSketchPU}
\end{figure}

This approach is limited since the additional corrections act on the four-momentum and not on the jet shape or substructure. To overcome this limitation, an alternative technique for PU mitigation, pileup per particle identification (PUPPI), is introduced \citep{Puppi}. It calculates, event by event, a probability that each particle originates from the leading primary vertex and scales the energy of these particles based on that probability (see \Cref{fig:LumiPileUpPuppiWeight}). As a consequence, objects clustered from hadrons, such as jets, \ptmiss, and lepton isolation, are expected to be less susceptible to PU when PUPPI is used. In \Cref{fig:PileupSketchPU}, a schematic representation of CHS and PUPPI is shown.

\begin{figure}[htb]
\centering
\subfigure{\label{fig:EffPuppi}}{\includegraphics[height=0.41\textwidth]{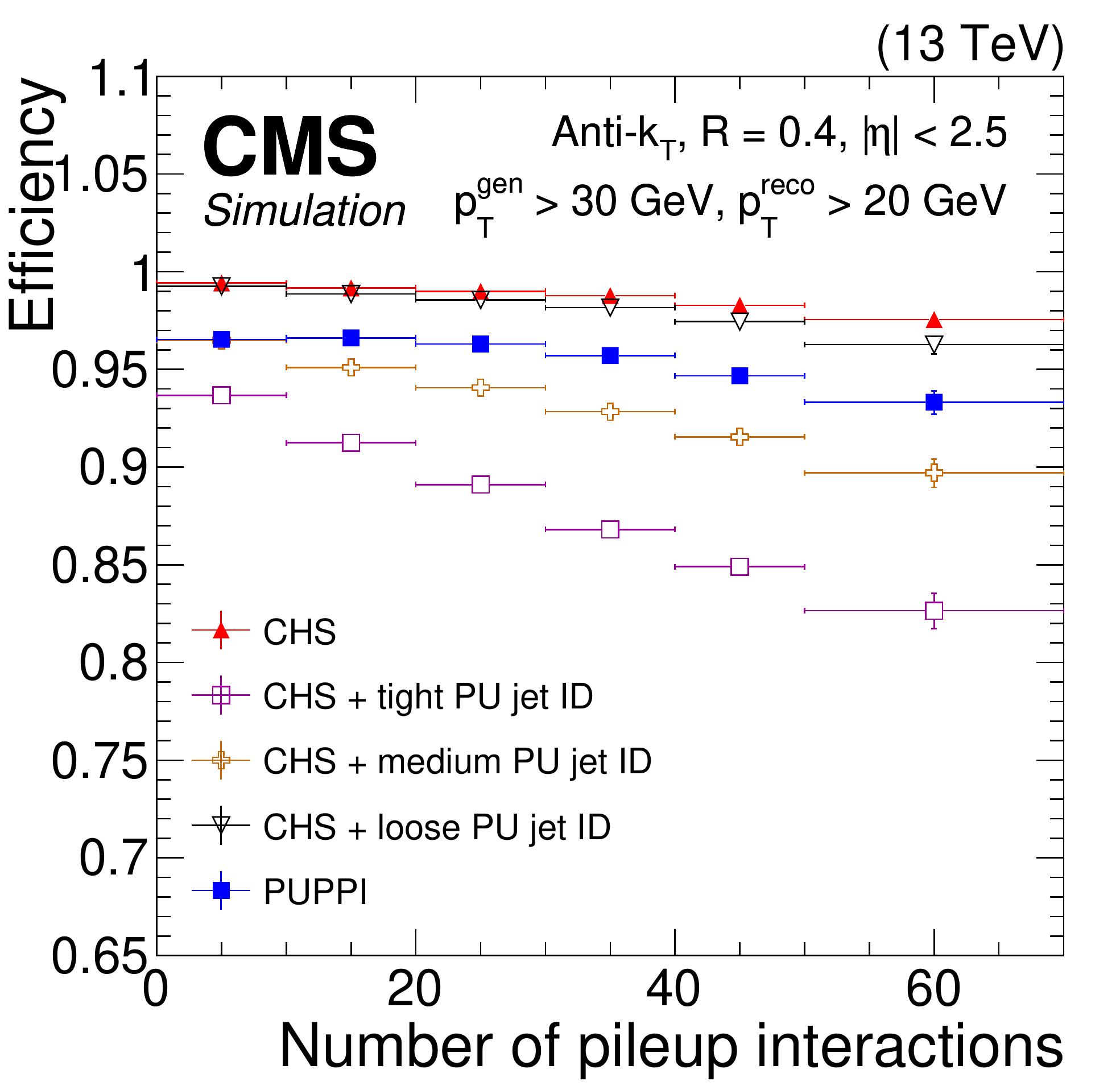}}
\subfigure{\label{fig:PurityPuppi2}}{\includegraphics[height=0.41\textwidth]{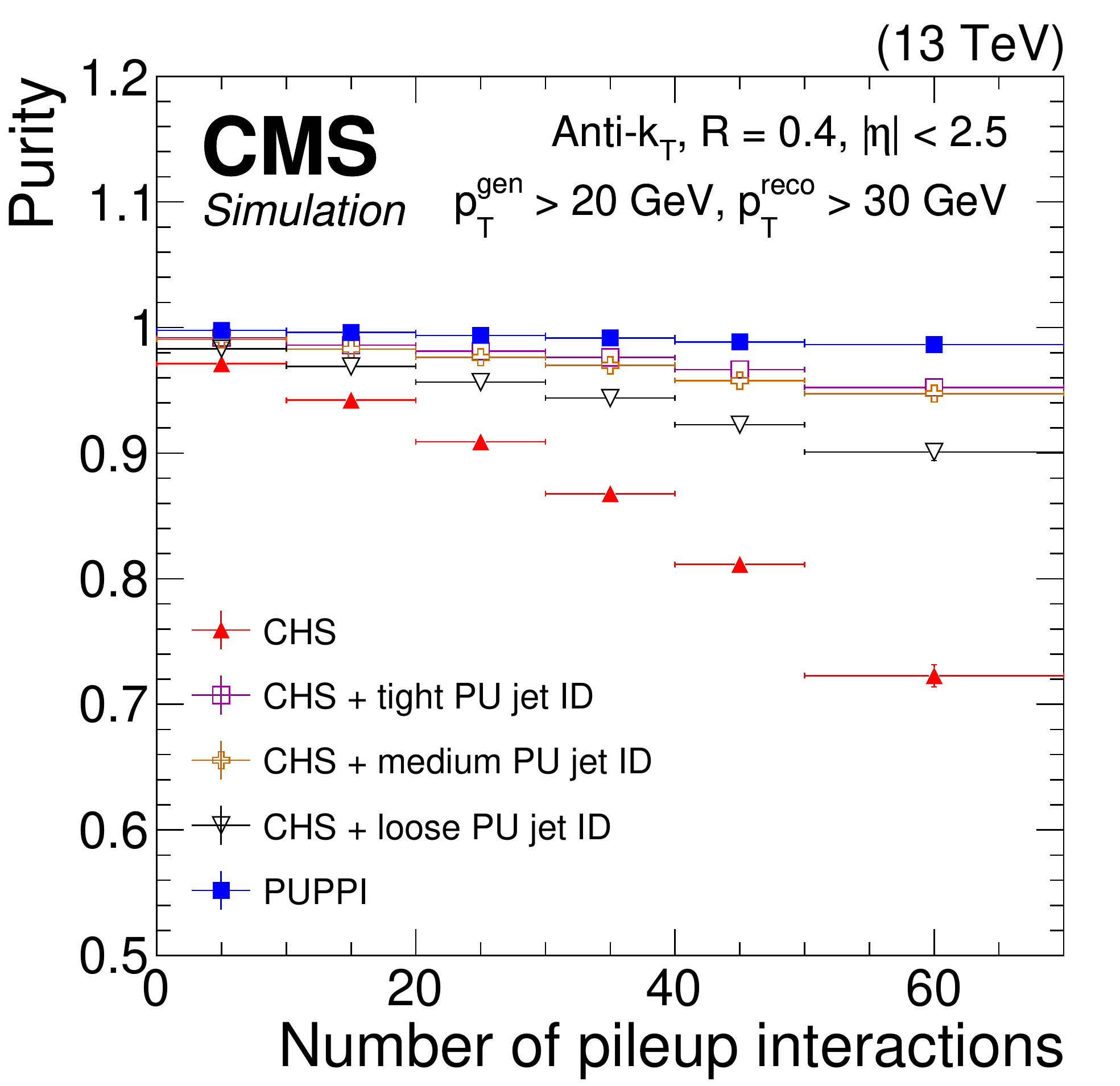}}
\caption{Efficiency and purity as a function of the number of interactions for PUPPI, CHS and CHS+PU jet ID at different working points \citep{Puppi}.}
\label{fig:PuppiEffPurity}
\end{figure}

These techniques are complementary, as highlighted in \Cref{fig:PuppiEffPurity}. Inside the tracker acceptance, PUPPI has a good performance in both efficiency and purity, defined as the fraction of reconstruction-level jets with $\pt\geq 30 \GeV$ that match within  $\Delta R\leq 0.4$ with a particle-level jet with $\pt\geq 20 \GeV$. In contrast, for CHS, even though  the efficiency is nearly close to 100\%, the purity is significantly reduced at high pileup. To improve the purity, but at the cost of a reduction in efficiency, one can apply the pileup jet ID, a boosted decision tree based technique to identify low-\pt jets coming from PU \citep{PUIDNotes}. At high values of $|\eta|$, the purity drops more rapidly in all the cases, and, even though PUPPI performs better than CHS only, the usage of pileup ID on top of CHS improves the performances compared to PUPPI. 

\section{Jet energy calibration}

\begin{figure}[htb]
\centering
\subfigure{\label{fig:JECL1offset}}{\includegraphics[height=0.41\textwidth]{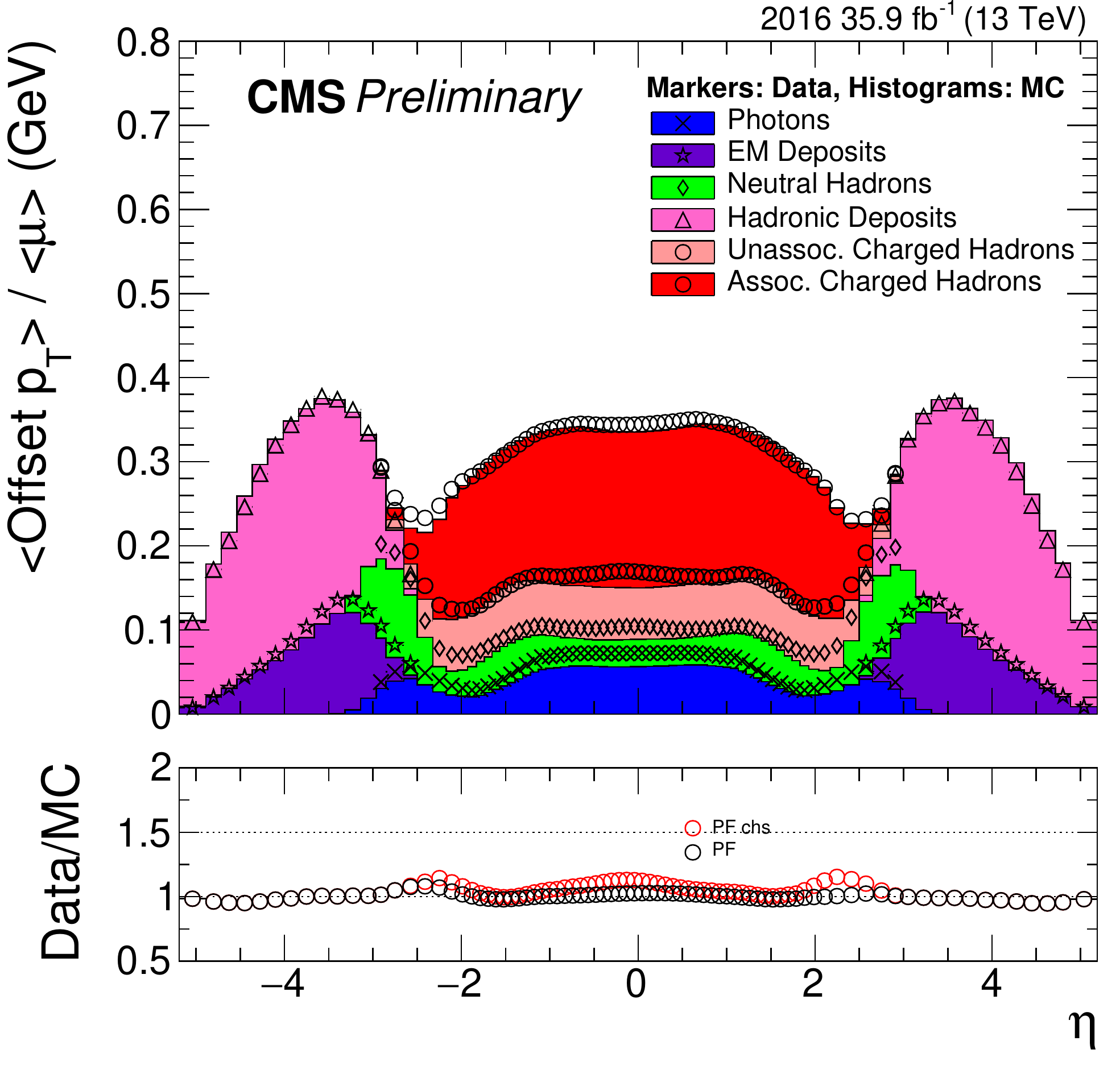}}
\subfigure{\label{fig:JECL2Relative}}{\includegraphics[height=0.41\textwidth]{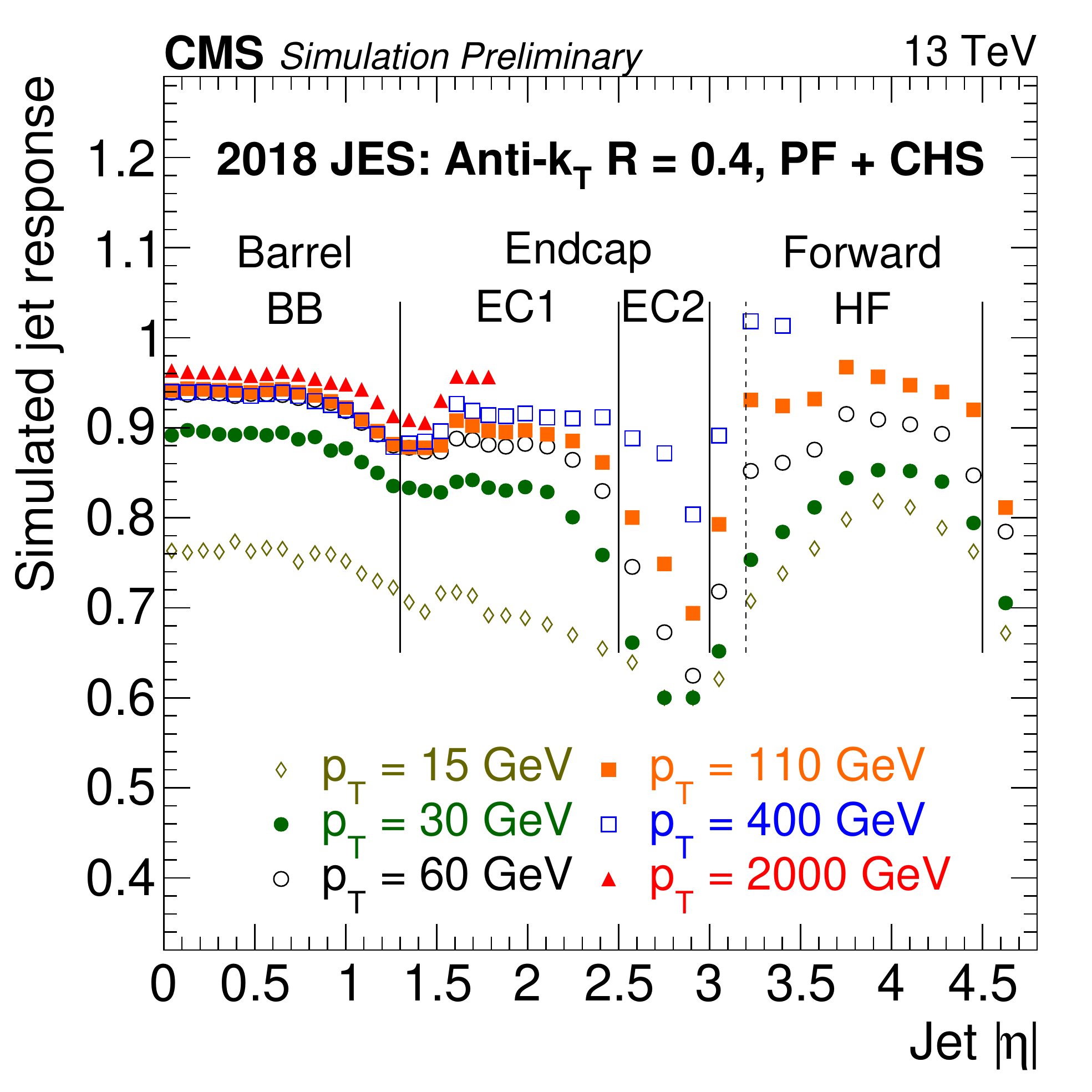}}
\caption{Left: Data-to-simulation comparison for average offset in \pt per pileup interaction, calculated for each type of PF candidate. Right: Jet response corrections derived as a function of $|\eta|$ and $\pt$. Changes in performance at high $|\eta|$ and low \pt are caused by detector acceptance. Taken from \citep{JERCNotes}.}
\label{fig:JEC}
\end{figure}

For jet energy calibration, CMS uses a factorised approach. The first step consists of PU subtraction. It is a simulation-based correction that aims to remove the average energy offset coming from PU, monitored for each type of PF candidate (see \Cref{fig:JEC}). The calibration procedure heavily relies on the second step, which corresponds to the calibration of the jet energy response (see \Cref{fig:JEC}). Here, the primary goal is to account for detector effects that introduce a discrepancy in the measured energy between the particle-level jets and the reconstructed ones.

Residual differences between data and simulation are then corrected in two steps: an $|\eta|$-dependent correction, to correct the different response of each sub-detector with respect to the central, better-calibrated part of the detector; a \pt-dependent correction, mitigating the absolute scale difference in the central region of the detector. 

\begin{figure}[htb]
\centering
\subfigure{\label{fig:JECL3Res}}{\includegraphics[height=0.4\textwidth]{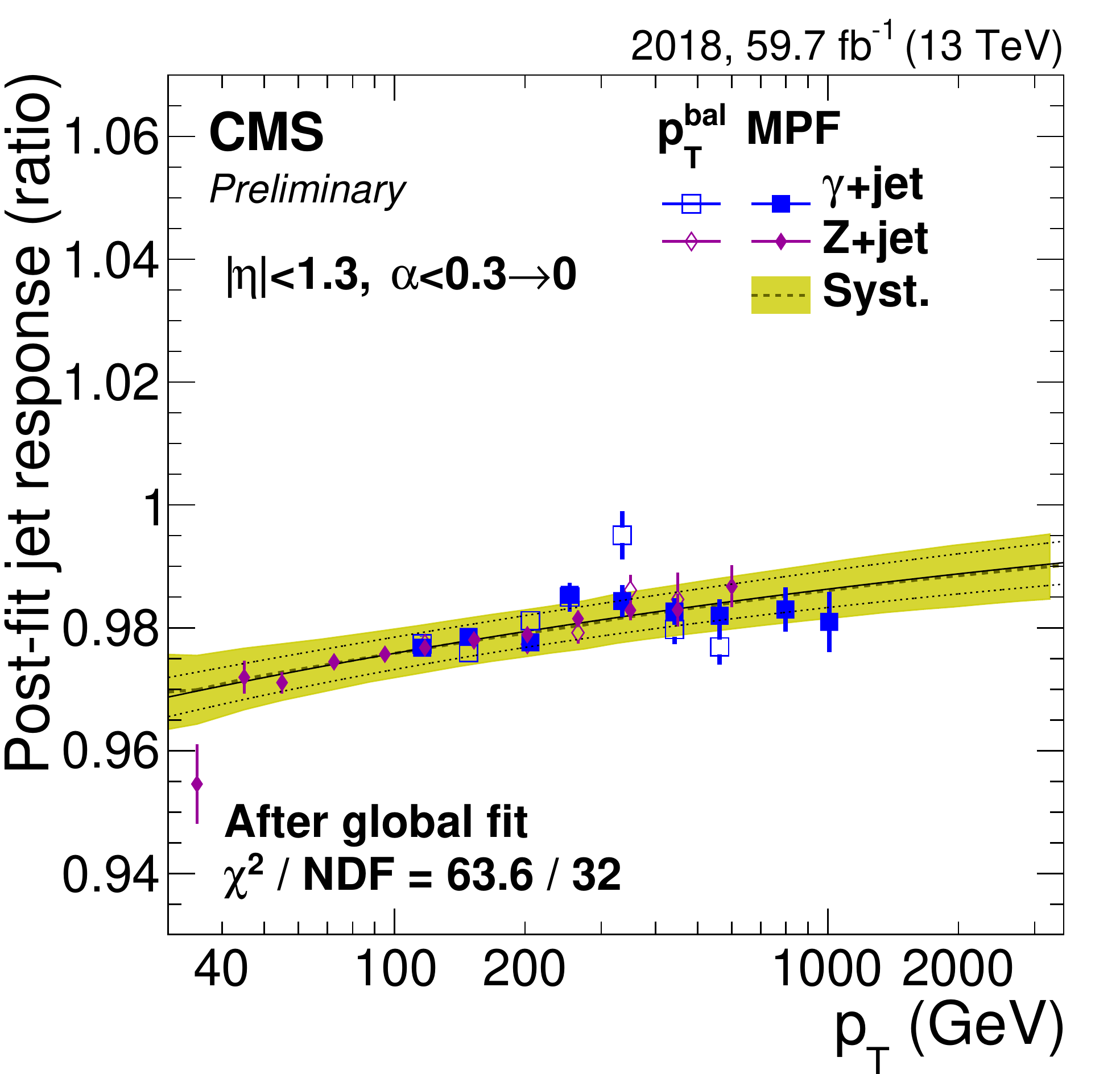}}
\subfigure{\label{fig:JER}}{\includegraphics[height=0.4\textwidth]{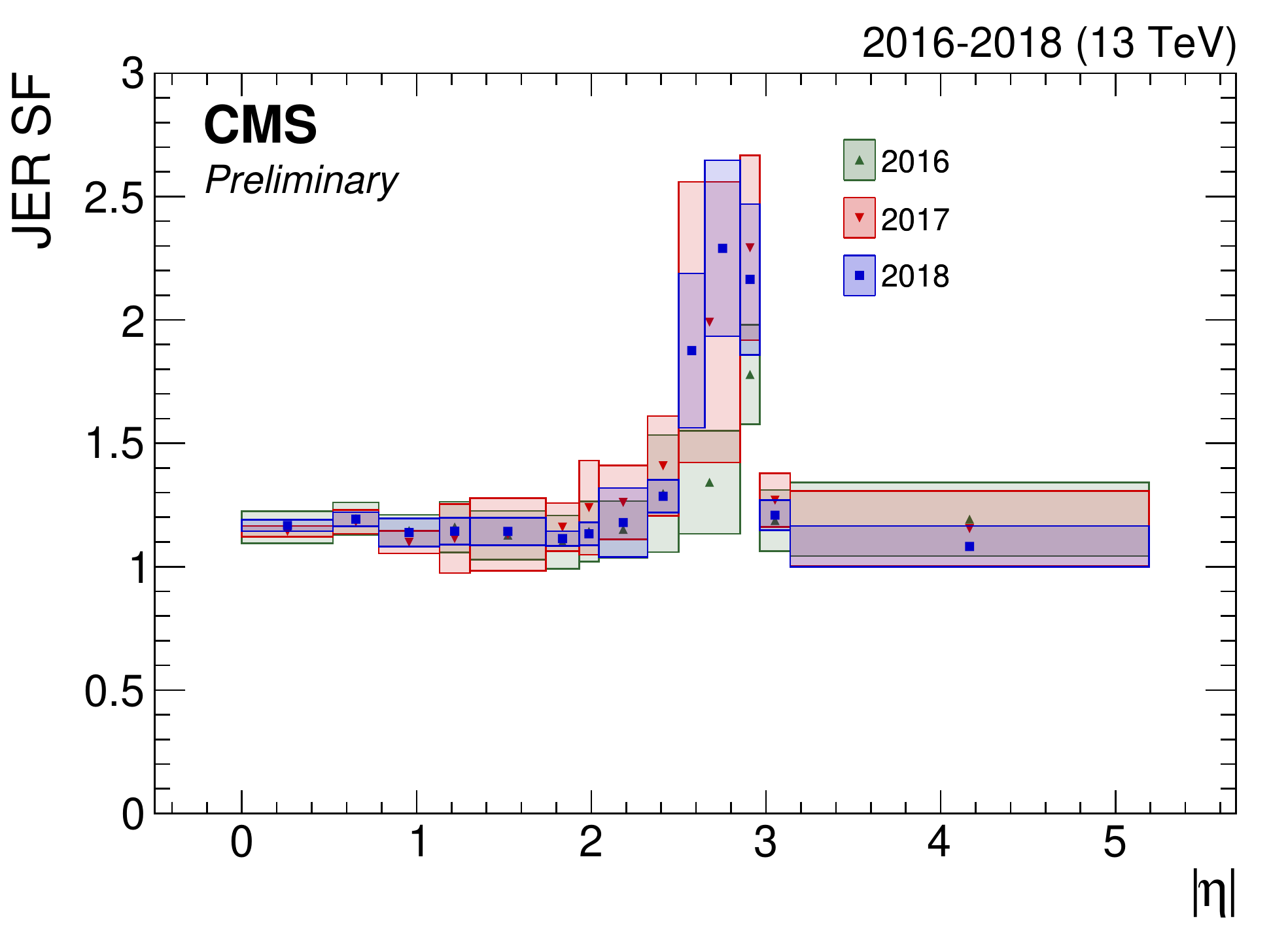}}
\caption{ Left: Data-to-simulation \pt-dependent residual corrections for the jet response. Right: $|\eta|$-dependent data-to-simulation jet energy resolution scale factors. Taken from \citep{JERCNotes}.}
\label{fig:JECUncJER}
\end{figure}

Several techniques involving precisely calibrated reference objects ( $Z\to\mu\mu$, $Z\to ee$ and $\gamma$) are used in a global fit to reach a precision at a percent level, as shown in  \Cref{fig:JECUncJER}.
After the jet energy scale has been corrected, the jet energy resolution (JER) in simulation needs to be adjusted to match the jet resolution in data. Therefore, scale factors (SFs) are applied in simulation to broaden the detector response distribution (see \Cref{fig:JECUncJER}). 
In the $\eta\in[2.5,3]$, the imperfect calibration of overlapping sub-detectors leads to larger SFs.

\section{Missing transverse momentum}

\begin{figure}[htb]
\centering
\subfigure{\label{fig:METfilters}}{\includegraphics[height=0.40\textwidth]{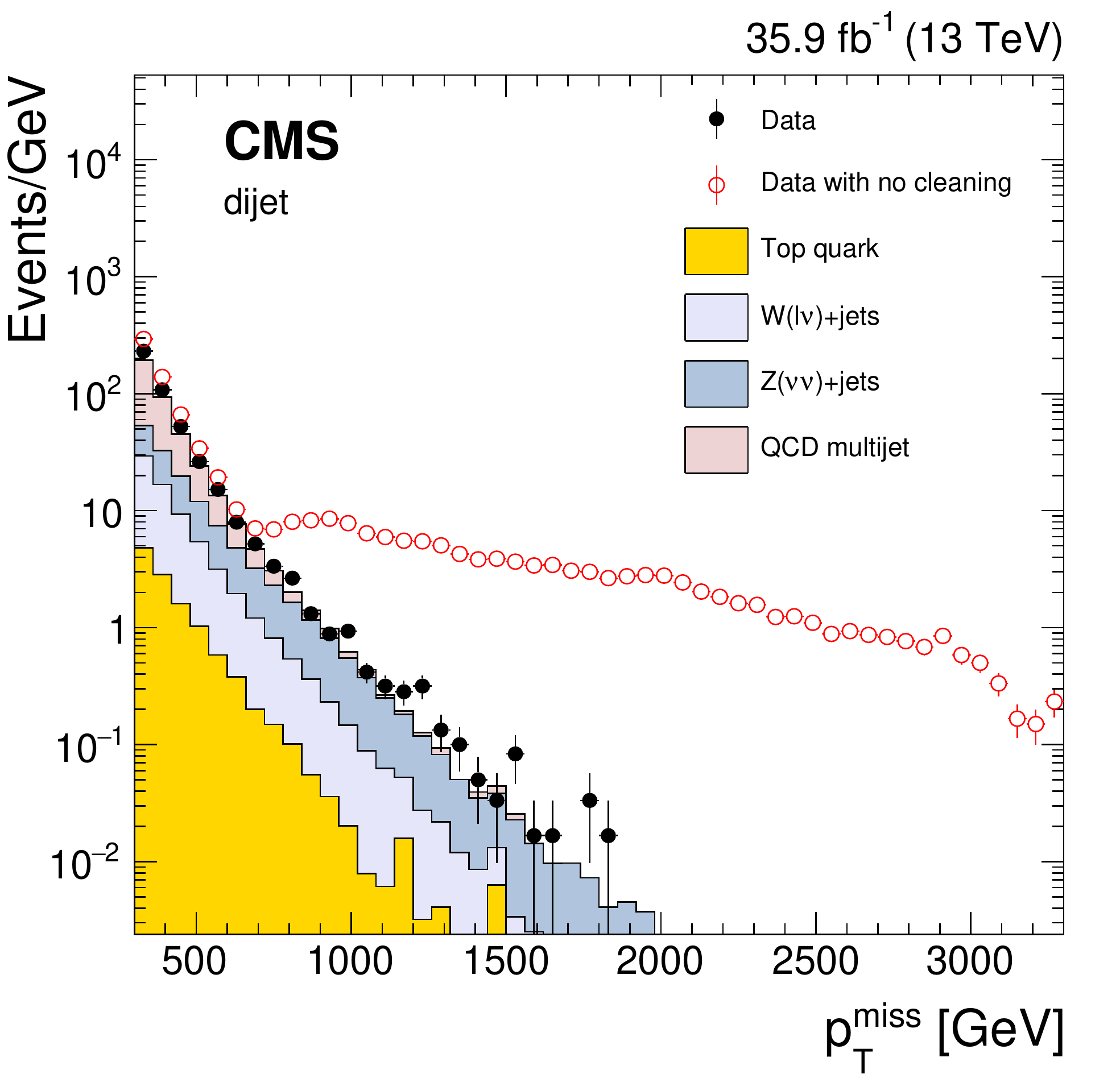}}
\subfigure{\label{fig:PuppiMETRes}}{\includegraphics[height=0.40\textwidth]{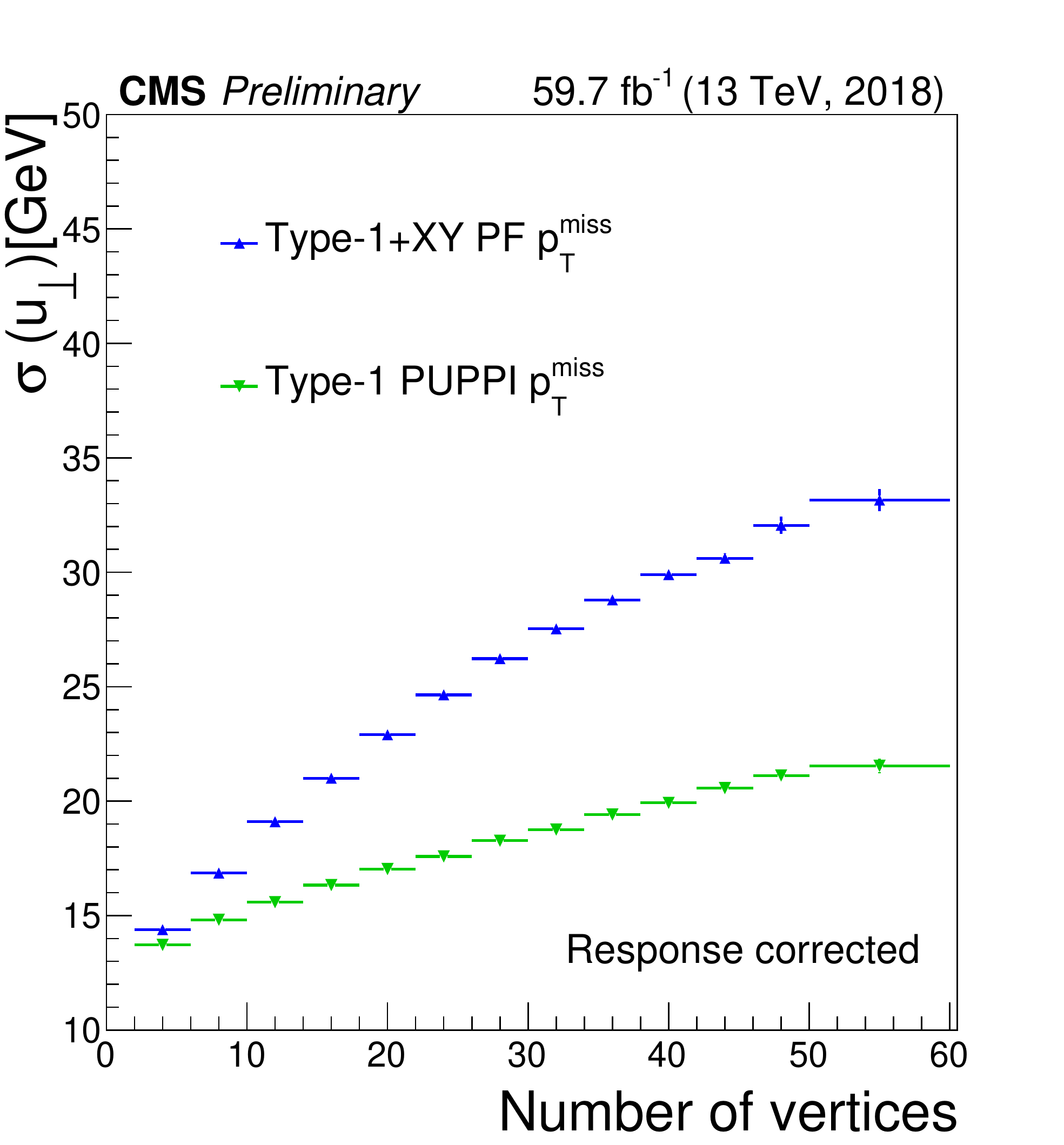}}
\caption{ Left: \ptmiss distribution with the event filtering algorithms applied on a dijet selection \citep{MET}. Right: The resolution of the hadronic recoil in $\gamma$+jets events as a function of the number of vertices for different PU suppression algorithms. Results for Run 2 are shown in \citep{METNotes}.
}
\label{fig:PerfMET}
\end{figure}

Missing transverse momentum arises from invisible particles, either from standard model neutrinos or new physics, not interacting with the detector. It is estimated from the momentum conservation in the transverse plane using the visible part of the event, which, if miscalibrated, can lead to an inaccurate estimation of \ptmiss \citep{MET}.
Furthermore, anomalous high-\ptmiss events can appear from detector effects. \Cref{fig:PerfMET} shows the effectiveness of several mechanisms adopted to suppress these spurious events in data.
The \ptmiss response and resolution can be assessed in events either with an isolated photon or with a Z boson decaying to a pair of electrons or muons, balancing the momentum of the vector boson to the hadronic recoil system ($\vec{u_T}$), where no genuine \ptmiss is expected.
The performance of the PUPPI algorithm on \ptmiss is presented in \Cref{fig:PerfMET}. A direct improvement in the resolution of transverse mass in W+jets events, where genuine \ptmiss is expected, is observed \citep{MET}. Also in this case, PUPPI shows better stability against PU, and the gain at the expected PU for Run 3 is significant.

\section{Summary}
Several techniques for jet and \MET reconstruction  and calibration adopted by the CMS Collaboration during Run 2 are presented.
The improved performance of pileup mitigation algorithms developed in this period will allow coping with the harsher conditions in the upcoming Run 3, during which PUPPI will be used as the default algorithm for pileup suppression.

\bibliographystyle{JHEP}
\bibliography{bibliography}

\providecommand{\href}[2]{#2}\begingroup\raggedright\begin{thebibliography}{1}

\bibitem{Collaboration_2008}
{CMS Collaboration}, \emph{The {CMS} experiment at the {CERN} {LHC}},
  \href{https://doi.org/10.1088/1748-0221/3/08/s08004}{\emph{Journal of
  Instrumentation} {\bfseries 3} (2008) S08004}
  https://doi.org/10.1088%2F1748-0221%2F3%2F08%2Fs08004.

\bibitem{PF}
{CMS Collaboration}, \emph{Particle-flow reconstruction and global event
  description with the cms detector},
  \href{https://doi.org/10.1088/1748-0221/12/10/p10003}{\emph{Journal of
  Instrumentation} {\bfseries 12} (2017) P10003–P10003}
  http://dx.doi.org/10.1088/1748-0221/12/10/P10003.

\bibitem{Puppi}
{CMS Collaboration}, \emph{Pileup mitigation at {CMS} in 13 {TeV} data},
  \href{https://doi.org/10.1088/1748-0221/15/09/p09018}{\emph{JINST} {\bfseries
  15} (2020) P09018} [\href{https://arxiv.org/abs/arxiv/2003.00503}{{\ttfamily
  arxiv:2003.00503}}].

\bibitem{PUIDNotes}
{CMS Collaboration}, \emph{{Performance of the pile up jet identification in
  CMS for Run 2}},  https://cds.cern.ch/record/2715906.

\bibitem{JERCNotes}
{CMS Collaboration}, \emph{{Jet energy scale and resolution performance with 13
  TeV data collected by CMS in 2016-2018}},
  https://cds.cern.ch/record/2715872.

\bibitem{MET}
{CMS Collaboration}, \emph{{Performance of missing transverse momentum
  reconstruction in proton-proton collisions at $\sqrt{s} =$ 13 TeV using the
  CMS detector}},
  \href{https://doi.org/10.1088/1748-0221/14/07/P07004}{\emph{JINST} {\bfseries
  14} (2019) P07004} [\href{https://arxiv.org/abs/1903.06078}{{\ttfamily
  1903.06078}}].

\bibitem{METNotes}
{CMS Collaboration}, \emph{{Performance of missing transverse momentum
  reconstruction in events containing a photon and jets collected by CMS during
  proton-proton collisions at $\sqrt{s}$ = 13 TeV in 2018}},
  https://cds.cern.ch/record/2723010.

\end{thebibliography}\endgroup

\end{document}